\input{aipcheck}
\documentclass[final]{aipproc}
\layoutstyle{8x11single}
\begin{document}

\title{Accretion History of Subhalo Population now and then}   

\classification{\texttt{98.80-k}}


\author{Carlo Giocoli}{address={Zentrum f\"ur Astronomie, ITA, 
Universit\"at Heidelberg, Albert-Ueberle-Str. 2, 69120 Heidelberg, 
              Germany},
email={cgiocoli@ita.uni-heidelberg.de},}

\begin{abstract}
In  the  standard model  of  structure  formation  galaxies reside  in
virialized   dark  matter   haloes  which   extend  much   beyond  the
observational  radius of  the central  system.  The  dark  matter halo
formation  process is  hierarchical,  small systems  collapse at  high
redshift and then merge together  forming larger ones. In this work we
study  the   mass  assembly  history  of  host   haloes  at  different
observation  redshifts and  the mass  function of  accreted satellites
(haloes that  merge directly  on the main  halo progenitor).   We show
that the satellite mass function is universal, both independent on the
host halo mass and  observation redshift.  The satellite mass function
also turn out to be universal once only satellites before or after the
host halo formation  redshift (time at which the  main halo progenitor
assembles half  of its final mass)  are considered.  We  show that the
normalizations of these distributions are directly related to the main
halo  progenitor mass  distributions before  and after  its formation,
while  their  slope  and  the  exponential high  mass  cut-off  remain
unchanged.
\end{abstract}
\maketitle
\section{Introduction}
Understanding the  structure formation process is  a fundamental topic
in modern cosmology.  In the current $\mathrm{\Lambda}$CDM concordance
cosmology, the  matter density  of the Universe  is dominated  by cold
dark  matter (CDM),  whose  gravitational evolution  gives  rise to  a
population of virialized  dark matter haloes spanning a  wide range of
masses \cite{sheth99}. Numerical simulations of structure formation in
a  CDM  universe predict  that  these  dark  matter haloes  contain  a
population of subhaloes, which are  the remnants of haloes accreted by
the  host,   and  which  are   eroded  by  the  combined   effects  of
gravitational heating and tidal stripping in the potential well of the
main halo.

Understanding the evolution of  the subhalo mass function, as a function
of  cosmology,  redshift,  and  host  halo  mass,  is  of  fundamental
importance,  with  numerous  applications.   For  one,  subhaloes  are
believed  to  host satellite  galaxies,  which  can  thus be  used  as
luminous tracers of the subhalo population. In particular, linking the
observed abundances of satellite galaxies to the expected abundance of
subhaloes,  provides  useful  insights  into  the  physics  of  galaxy
formation  \cite{moore99,bullock00,somerville02}. Studies  along these
lines indicate that galaxy  formation becomes extremely inefficient in
low mass haloes, and suggest that there may well be a large population
of       low      mass       subhaloes      with       no      optical
counterpart \cite{kravtsov04,stoehr02}.

In principle,  though, these truly 'dark' subhalos  may potentially be
detected via $\gamma$-ray emission  due to dark matter annihilation in
their                                                           central
cores \cite{stoehr03,bertone06,pieri08,giocoli08a,springel08,giocoli09}
or via  their impact on  the flux-ratio statistics  of multiply-lensed
quasars \cite{metcalf01,dalal02}.  Alternatively, these techniques may
be used  to constrain  the abundance of  subhaloes, which in  turn has
implications  for cosmological  parameters and/or  the nature  of dark
matter.

The evolution of  the subhalo mass function is  also of importance for
the survival  probability of disk  galaxies \cite{toth92,benson04} and
even has implications for direct detection experiments of dark matter.

Finally, understanding  the rate at  which dark matter  subhaloes lose
mass has  important implications  for their dynamical  friction times,
and       thus        for       the       merger        rates       of
galaxies \cite{benson02,zentner03,taylor04}.

Despite  significant  progress in  the  last  years,  there are  still
numerous issues  that are insufficiently understood. What  is the mass
function of haloes accreted onto  the main progenitor of a present day
host halo?   Does the  distribution depend on  the host halo  mass and
observation redshift? Answareing this  question is also very important
because,  while   the  mass  of  the  satellite   decrease  under  the
gravitational influence  of the host due to  gravitational heating and
tidal    stripping,    its     luminosity    is    thought    to    be
preserved \citet{vale06,vale08,li09}.   In this work  we address these
questions using high resolution  numerical simulations.  We trace back
the  evolution  of host  haloes  starting  from different  observation
redshifts, $z_0  = 0,\,0.5,\,1$ and $2$,  identifing progenitor haloes
directly accreted onto the main branch.

\section{Simulation and Post Processing}

Let us consider  the outputs of a cosmological  N-Body simulation of a
$\mathrm{\Lambda}$CDM  universe  in  a  periodic  cube  of  side  $100
h^{-1}$Mpc.  This  simulation  has  a  total number  of  particles  of
$400^3$,    with   an    individual   mass    of    $m_p=1.73   \times
10^{9}h^{-1}M_{\odot}$.  The  cosmological parameters are ($\Omega_m$,
$\sigma_8$,  $h$,  $\Omega_b h^2$)  =  (0.3,  0.9,  0.7 0.0196),  more
details about the GIF2 simulation and the post processing can be found
in
\cite{giocoli08b} and \cite{gao04}.

At  each simulation  snapshot  haloes have  been  identified with  the
spherical overdensity criterion.  Having estimated the density of each
particles and  sorted them by density,  we start from  the densest one
and grow a sphere of matter  around its center, and stop when the mean
density   within   the   sphere    first   fall   below   the   virial
value \cite{eke96}. At  this point we assign all  particles within the
sphere to the newly formed halo, and remove them from the global list.
We take  the centre of  the next halo  at the position of  the densest
particle  among the  remaining  ones  and grow  a  second sphere.   We
continue in this manner until  all particles are assigned to haloes or
to the field.

Starting from a given observation  redshift $z_0$ we build the merging
history tree,  for all  haloes\footnote{Haloes which exceed  more then
$10\%$ the  initial virial mass at  any $z>z_0$ are  not considered in
this    analysis.}    in    the   simulation    more    massive   then
$10^{11.5}h^{-1}M_{\odot}$,   using  the   halo   catalogues  at   all
snapshots,   separated  by   redshift   inteval  $\mathrm{d}z_i$,   as
follows. Considering  a halo  with virial mass  $M_{z_0}$ at  $z_0$ we
define   its    progenitors   at   the    previous   output   $z_1=z_0
+ \mathrm{d}z_1$ to be all haloes contributing with at least $50\%$ of
their particles  to the initial system.   Among them we  call the main
halo progenitor  the halo providing the largest  mass contribution. We
repeat  the  procedure, starting  now  at  $z_1$  and considering  the
progenitor  at $z_2  = z_1  +  \mathrm{d}z_2$ of  the $z_1$-main  halo
progenitor, and so backward in redshift.  In this merging history tree
we  term \emph{satellite} all  progenitors which,  at any  time, merge
directly on the main progenitor,  contributing at least with $50\%$ of
their particles to the initial system at $z=z_0$.

In  the hierarchical  growth of  the  parent host  haloes we  identify
$z_f$, the formation redshift of the halo, to be the redshift at which
the main halo progenitor assembles half of its final mass \cite{lc93}.
Because small systems collapse first and merge together we expect more
(less)   massive   haloes  to   have   a   lower  (higher)   formation
redshift \cite{giocoli07}.

\begin{figure}[h]
\centering
\includegraphics[width=.5\hsize]{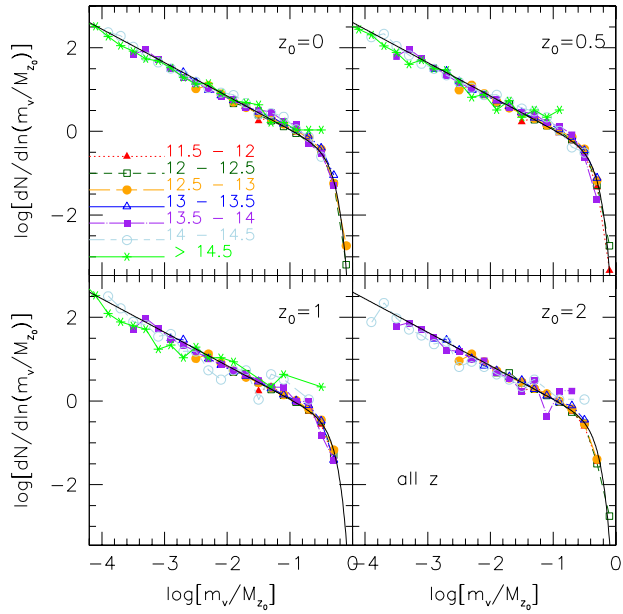}
\caption{\label{unevfig} Mass accreted in satellites 
(\emph{unevolved}  subhalo mass function)  by the  main branch  of the
tree at  all redshifts.  In each  panel we show  the results following
the tree starting from  different observation redshifts $z_0$ and host
halo  masses.   The  solid   lines  represent  the  fitting  function:
Equation \ref{uneveq}.}
\end{figure}

\subsection{Satellite Mass Function}

In Figure  \ref{unevfig} we show  the satellite mass function  at four
different observation redshifts $z_0=$0, 0.5, 1 and 2, the virial mass
of each satellite  $m_{\rm v}$ is expressed in units  of the host halo
initial mass $M_{z_0}$.  In each  panel different data points and line
types refer  to various  host halo masses.  From the figure  we notice
that all data points, at each observation redshift, are well fitted by
the following equation:
\begin{equation}
\frac{\mathrm{d}N}{\mathrm{d}\ln(m_v/M_{z_0})} = N_0 \, x^{-\alpha}
\mathrm{e}^{-6.283 x^3},\,\,\,x = \left| \frac{m_v}{\alpha M_{z_0}} \right|\,,
\label{uneveq}
\end{equation}
which is represented  by the solid curve. The  normalization $N_0$ and
the   slope   $\alpha$   are    respectively   equal   to   0.21   and
0.8 \cite{giocoli08b}. This  universal distribution has been predicted
by \cite{lc93}  using the  extendend-Press \& Schechter  formalism and
linear  scale-free  power   spectrum.   \cite{vandenbosch05}  found  a
different  slope and  normalization following  haloes from  $z_0=0$ to
high redshift using a Monte-Carlo  merger tree realization of halos in
a $\mathrm{\Lambda}$CDM cosmology However the first measurement of the
universality  of this  distribution in  numerical simulation  has been
done by \cite{giocoli08b} following haloes only again from $z_0=0$. In
this  work  we  show that  also  if  we  follow haloes  starting  from
different   observation  redshifts,   that  are   $z_0  \ge   0$,  the
distributions still  preserve their universality  with identical slope
and normalization.

\begin{figure}[!h]
\centering
\includegraphics[width=.45\hsize]{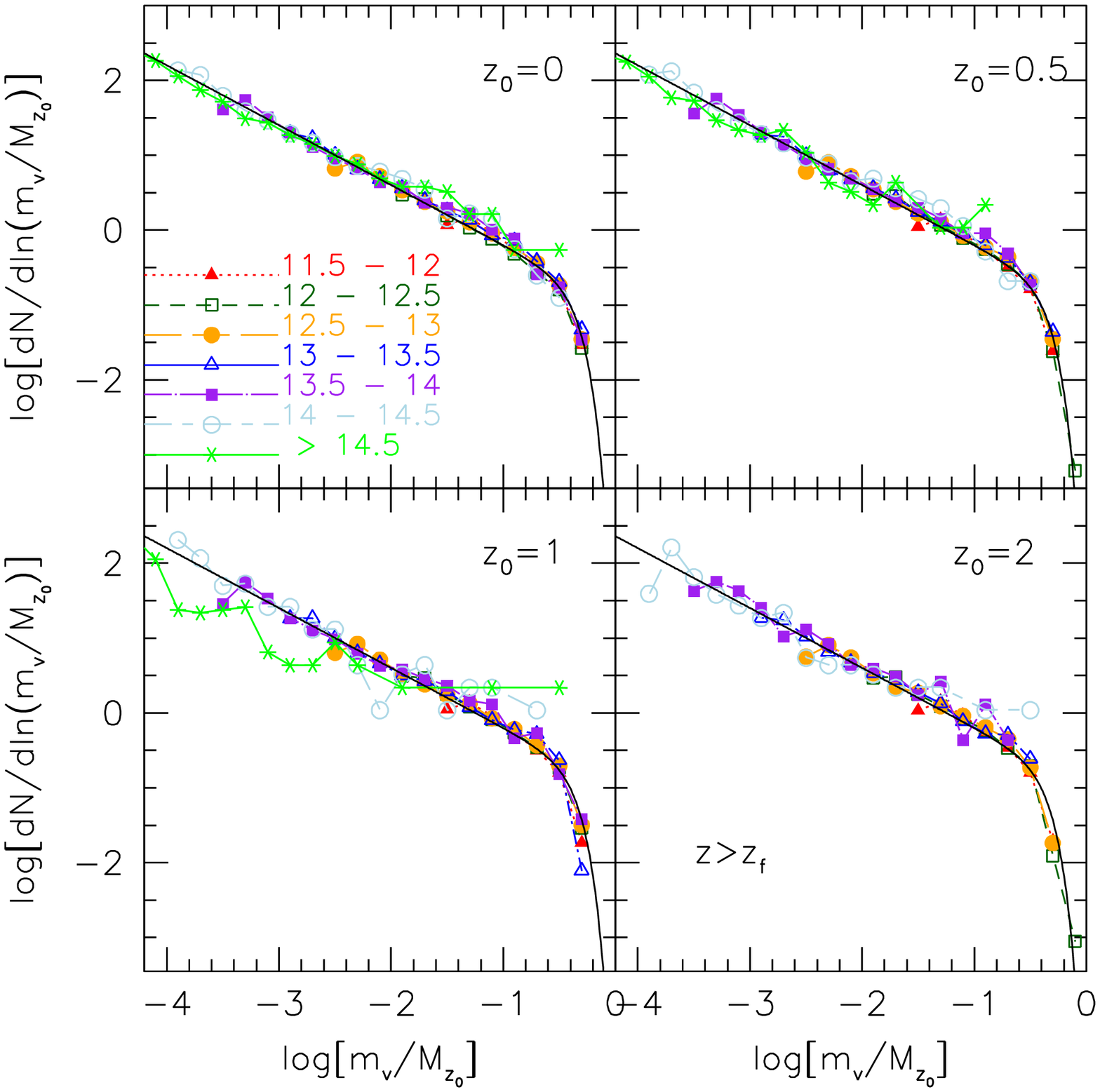}
\includegraphics[width=.45\hsize]{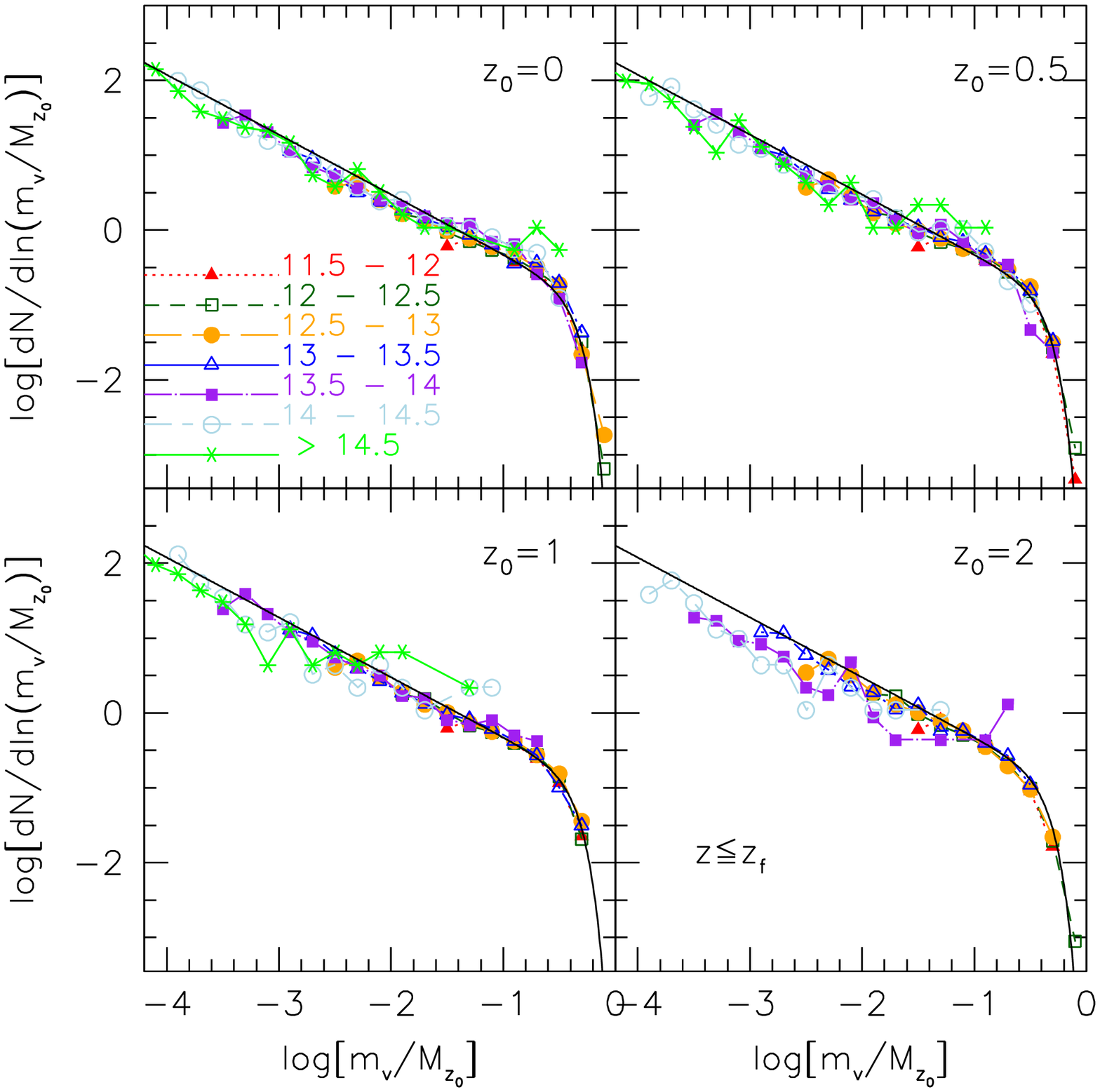}
\caption{\label{unevbazf} Satellite mass function accreted before
(left) and  after (right) the host halo  formation redshift. Different
data points and line types refer to varius final host halo masses. The
solid curve represent equation  \ref{uneveq} with the correct rescaled
normalizations.}
\end{figure}

In  Figure \ref{unevbazf}  we  show the  satellite  mass function  for
different  initial  host halo  masses  and  observation redshift,  but
considering  satellite progenitors  accreted before  (left)  and after
(right)    the     formation    redshift    of     the    main    halo
progenitor. Equation \ref{uneveq} also  in these cases fits quite well
the data point distributions but with different normalizations that we
term $N_{0,\rm b}$ (for $z>z_f$) and $N_{0,\rm a}$ (for $z \le z_f$).

\subsubsection{Following the Main Branch}

In  order to  estimate the  normalization of  the distribution  of the
satellite mass function before and after the formation redshift of the
host  halo  we need  to  compute the  distribution  of  the main  halo
progenitor at $z_f$.   This is because, if $M_{z_0}$  is the host halo
mass  at $z_0$,  and all  the mass  is assembled  throught  merging of
progenitor  haloes at $z_1$,  requiring the  mass conservation  we can
write:
\[
 \sum_i m_{i,z_1} + M_{z_1} = M_{z_0}\,,
\]
where we  divided the sum into  the total mass in  satellite haloes plus
the  main halo  progenitor. At  this  point we  can also  do the  same
considering  the  progenitor  haloes   of  $M_{z_1}$  at  $z_2  =  z_2
+ \mathrm{d}z_2$, which gives:
\[
 \sum_i m_{i,z_1} + \sum_i m_{i,z_2} + M_{z_2} = M_{z_0}\,.
\]
Following the  main halo  progenitor back in  time until  it assembles
more then half of its initial mass we can write:
\begin{equation}
  \sum_{j=0}^{z_j\le   z_f}   \sum_i  m_{i,z_j}   +   \mu  M_{z_0}   =
  M_{z_0}\,, \label{cons1}
\end{equation}
where $\mu  = M_{z_f}/M_{z_0}$ represents the  initial main progenitor
mass fraction at $z_f$ and $\sum_j^{z_j \le z_f} \sum_i m_{i,z_j}$ the
total mass accreted in satellites for  $z \le z_f$. This first term on
the  left hand  side of  Equation \ref{cons1}  is directly  related to
$N_{0.a}$,  the normalization of  the satellite  mass function  in the
right panels of Figure \ref{unevbazf}.

The  analytic  estimate  of  $\mu$  can be  obtained  considering  the
distribution of the main halo progenitor mass soon after the formation
redshift,
\begin{equation}
p(\mu) \mathrm{d}\mu = \frac{2}{\pi} \sqrt{\frac{1-\mu}{2\mu-1}} 
\frac{\mathrm{d}\mu}{\mu^2}\,,\label{st04a}
\end{equation}
with  $1/2 \le  \mu  \le  1$, estimated  by  \cite{sheth04} using  the
extended-Press \& Schechter formalism.

\begin{figure}[!h]
\centering
\includegraphics[width=.5\hsize]{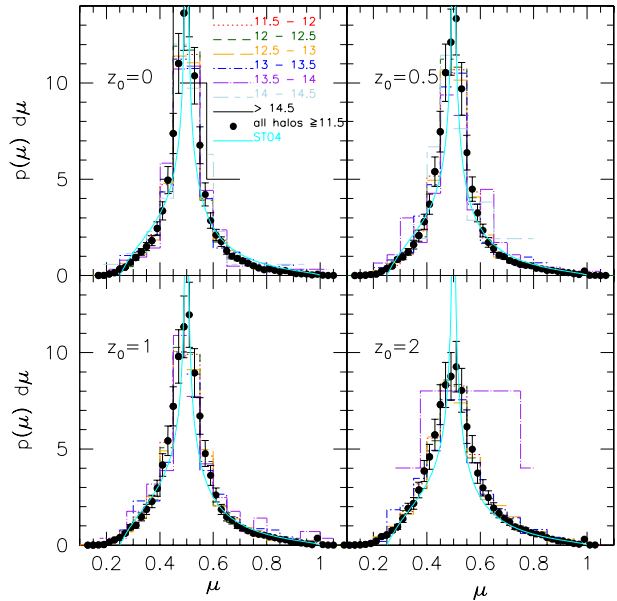}
\caption{\label{formfig} Formation mass distribution measured in 
the GIF2 simulation for different final halo mass bins and observation
redshifts with $1/2  \le \mu \le 1$. The  various line type histograms
show the result of different  final host halo masses. Considering that
the distribution does  not depend on $M_0$, we plot  all the halo more
massive  than  $10^{11.5}   M_{\odot}/h$  with  filled  circles.   The
corresponding error bars assume Poisson counts.  For $\mu \le 1/2$ the
mass distribution  just before  the formation is  shown. See  the main
text for more details.}
\end{figure}

Performing one  more step  along the  main branch of  the tree  we can
re-write  Equation \ref{cons1}  and  obtain the  main halo  progenitor
distribution soon before its formation:
\[
  \sum_{j=0}^{z_j\le  z_f} \sum_i  m_{i,z_j} +  \sum_{i} m_{i,z_{f+1}}
  + \mu' M_{z_0} = M_{z_0}\,,
\]
where $\mu'=M_{z_{f+1}}/M_{z_0}$  is the initial  main progenitor mass
fraction  assembled soon  before  the formation  redshift. The  $\mu'$
distribution  has also been  estimated by  \cite{sheth04} and  has the
following equation:
\begin{equation}
 p(\mu') \mathrm{d} \mu' = \frac{\mathrm{d}\mu' / \mu'^2}{\pi(1-\mu')}
\left( \sqrt{\frac{\mu'}{1-2 \mu'}} - \sqrt{1-2 \mu'}\right)\,,
\label{st04b}
\end{equation}
where $1/4 \le \mu' \le 1/2$. In this case the mean value of $\mu'$ is
related to the total mass accreted  in satellites for $z > z_f$, which
is proportional to $N_{0.b}$,  the normalization of the satellite mass
function in the left panels of Figure \ref{unevbazf}.

In  Figure  \ref{formfig}  we  show  the  main  halo  progenitor  mass
distribution   before   ($1/4   \le    \mu   \le   1/2$)   and   after
($1/2  \le  \mu  \le  1$)  the formation  redshift.   Histograms  with
different  line types  refer to  various  initial host  halo masses  at
corresponding observation redshifts $z_0$ in the GIF2 simulation, while
the data points  refer to all host haloes  considered with mass larger
than $10^{11.5}h^{-1}M_{\odot}$.   The two  solid lines in  each panel
represent  Equations  \ref{st04a}   and  \ref{st04b}  which  perfectly
reproduce the simulation results. Mass conservation implies that:
\[
 N_0 = N_{0,\rm a} + N_{0,\rm b}\,,
\]
and for the  mean main halo progenitor mass  assembled soon before and
after the formation redshift:
\[
 1 = \bar{\mu}' + \bar{\mu} \,.
\]
The normalizations $N_{0,\rm a}$ and $N_{0,\rm b}$ have been estimated
measuring the mean halo progenitor  mass, for all haloes starting from
every  observation  redshift,  soon  before and  after  the  formation
redshift:
\begin{eqnarray}
 N_{0,\rm b} &=& \bar{\mu} N_0 = (0.583 \pm 0.09) N_0 \nonumber \\ 
 N_{0,\rm b} &=& \bar{\mu}' N_0 = (0.432 \pm 0.06) N_0 \nonumber\,,
\end{eqnarray}
absolutely compatible within the  error bar with the mass conservation
equations above.  The small deviation from unity of the sum $\bar{\mu}
+ \bar{\mu}'$  is due  to the smooth  particle component (mass  not in
haloes) accreted  in numerical simulations  along the main  branch. In
the  panels in Figure  \ref{unevbazf} the  solid curves  represent the
satellite  mass  function (Equation  \ref{uneveq})  with the  obtained
rescaled  normalizations  that  fit  quite well  the  respective  data
points.

\section{Summary Remarks}
The standar scenario of  structure formation predicts that dark matter
haloes grow as consequence of repeated merges events. Central galaxies
reside in  the main halo  progenitor center, while  satellite galaxies
are  in  subhaloes.  These  represent  the  core  of surviving  haloes
accreted by the  main halo progenitor during its  cosmic evolution. In
these pages we  have measured the satellite mass  function accreted by
the main  branch of the merger  tree of various  host haloes, starting
from  different  observation  redshift  $z_0$ in  a  numerical  N-Body
simulation.  The  satellite mass function  turns out to  be universal:
both independent of the host halo mass and redshift.  The shape of the
distribution  remains unchanged  also when  satellite  haloes accreted
before and after the formation  redshift are considered.  In this case
the normalizations  can be obtained studying the  main halo progenitor
distribution soon before and after its formation.

\begin{theacknowledgments}
Thanks to Ravi  K. Sheth, Giuseppe Tormen and  my fellow travellers in
Paris for useful and  stimolating discussions, thanks also to Matthias
Bartelmann for having careflully read the manuscript.
\end{theacknowledgments}

\bibliographystyle{aipproc}

\begin{thebibliography}{}

\bibitem{sheth99} Sheth, R.~K., \& Tormen, G.\ 1999, MNRAS, 308, 119 

\bibitem{moore99} 
Moore, B., Ghigna, S., Governato, F., Lake, G., Quinn, 
T., Stadel, J., \& Tozzi, P.\ 1999, ApJL, 524, L19.

\bibitem{bullock00} Bullock, J.~S., 
Kravtsov, A.~V., \& Weinberg, D.~H.\ 2000, ApJ, 539, 517 

\bibitem{somerville02} Somerville, R.~S.\ 2002, 
ApJL, 572, L23 

\bibitem{kravtsov04} Kravtsov, A.~V., 
Berlind,  A.~A., Wechsler,  R.~H., Klypin,  A.~A.,  Gottl{\"o}ber, S.,
Allgood, B., \& Primack, J.~R.\ 2004, ApJ, 609, 35

\bibitem{stoehr02} Stoehr, F., White, 
S.~D.~M., Tormen, G., \& Springel, V.\ 2002, MNRAS, 335, L84 

\bibitem{stoehr03} Stoehr, F., White, 
S.~D.~M., Springel, V., Tormen, G.,  \& Yoshida, N.\ 2003, MNRAS, 345,
1313

\bibitem{bertone06} Bertone, G.\ 2006, PhRvD, 73, 103519 

\bibitem{pieri08} Pieri, L., Bertone, G., \& Branchini, E.\ 2008, MNRAS, 384, 1627

\bibitem{giocoli08a} Giocoli, C., Pieri, L., \& Tormen, G.\ 2008, MNRAS, 387, 689 

\bibitem{giocoli09} Giocoli, C., Pieri, L., 
Tormen, G., \& Moreno, J.\ 2009, MNRAS, 395, 1620

\bibitem{springel08} Springel, V., et al.\ 2008, MNRAS, 391, 1685 

\bibitem{metcalf01} Metcalf, R.~B., \& Madau, P.\ 2001, ApJ, 563, 9

\bibitem{dalal02} Dalal, N., \& Kochanek, C.~S.\ 2002, ApJ, 572, 25 

\bibitem{toth92} Toth, G., \& Ostriker, J.~P.\ 1992, ApJ, 389, 5 

\bibitem{benson04} Benson, A.~J., Lacey, 
C.~G., Frenk, C.~S., Baugh, C.~M., \& Cole, S.\ 2004, MNRAS, 351, 1215 

\bibitem{benson02} Benson, A.~J., Ellis, R.~S., \& Menanteau, F.\ 2002, MNRAS, 336, 564 

\bibitem{zentner03} Zentner, A.~R., \& Bullock, J.~S.\ 2003, ApJ, 598, 49 

\bibitem{taylor04} Taylor, J.~E., \& Babul, A.\ 2004, MNRAS, 348, 811 

\bibitem{vale06} Vale, A., \& Ostriker, J.~P.\ 2006, MNRAS, 371, 1173 

\bibitem{vale08} Vale, A., \& Ostriker, J.~P.\ 2008, MNRAS, 383, 355 

\bibitem{li09} Li, R., Mo, H.~J., Fan, Z., Cacciato, M., van den Bosch, F.~C., Yang, X., 
\& More, S.\ 2009, MNRAS, 394, 1016 

\bibitem{giocoli08b} Giocoli, C., Tormen, 
G., \& van den Bosch, F.~C.\ 2008, MNRAS, 386, 2135

\bibitem{gao04} Gao, L., White, S.~D.~M., 
Jenkins, A., Stoehr, F., \& Springel, V.\ 2004, MNRAS, 355, 819 

\bibitem{eke96} Eke, V.~R., Cole, S., \& Frenk, C.~S.\ 1996, MNRAS, 282, 263 

\bibitem{lc93} Lacey, C., \& Cole, S.\ 1993, MNRAS, 262, 627 

\bibitem{giocoli07} Giocoli, C., Moreno, J., Sheth, R.~K., \& Tormen, G.\ 2007, 
MNRAS, 376, 977 

\bibitem{vandenbosch05} van den Bosch, F.~C., Tormen, G., \& Giocoli, C.\ 
2005, MNRAS, 359, 1029 

\bibitem{sheth04} Sheth, R.~K., \& Tormen, G.\ 2004, MNRAS, 349, 1464 


\end{thebibliography}

\end{document}